# `Gene-R1`: Reasoning with Data-Augmented Lightweight LLMs for Gene Set Analysis[*]

Zhizheng Wang[†], Yifan Yang[†], Qiao Jin, Zhiyong Lu[‡]

*Division of Intramural Research (DIR), National Library of Medicine (NLM),
National Institutes of Health (NIH), Bethesda, MD 20894, USA
Email: {zhizheng.wang; yifan.yang3; qiao.jin; zhiyong.lu}@nih.gov*

The gene set analysis (GSA) is a foundational approach for uncovering the molecular functions associated with a group of genes. Recently, LLM-powered methods have emerged to annotate gene sets with biological functions together with coherent explanatory insights. However, existing studies primarily focus on proprietary models, which have been shown to outperform their open-source counterparts despite concerns over cost and data privacy. Furthermore, no research has investigated the application of advanced reasoning strategies to the GSA task. To address this gap, we introduce **Gene-R1**, a data-augmented learning framework that equips lightweight and open-source LLMs with step-by-step reasoning capabilities tailored to GSA. Experiments on 1,508 in-distribution gene sets demonstrate that Gene-R1 achieves substantial performance gains, matching commercial LLMs. On 106 out-of-distribution gene sets, Gene-R1 performs comparably to both commercial and large-scale LLMs, exhibiting robust generalizability across diverse gene sources.

*Keywords:* Gene Set Analysis; Data Augmentation; Fine-tuning; Reasoning; LLM

## 1. Introduction

Gene set analysis (GSA) is a foundational approach for revealing the molecular functions associated with groups of genes involved in physiological processes, healthcare, and disease[1,2]. By identifying the biological functions enriched in gene sets, GSA provides critical insights for elucidating disease mechanisms and discovering therapeutic targets[3,4]. Such mechanistic insights would greatly advance our understanding of functional genomics.

Over its development, GSA has progressed through two notable methodological paradigms: classical functional enrichment analysis (shown in **Fig.1 (a)**) and emerging solutions based on large language models (LLMs) (shown in **Fig.1 (b)**). The traditional methods[5,6] typically compare gene sets against predefined categories in manually curated databases such as Gene Ontology (GO)[7] and Molecular Signatures Database (MSigDB)[8] to identify functions that are statistically significantly enriched. The LLM-powered approaches aim to generate biological functional annotations and coherent explanatory narratives for gene sets through instruction learning[2,9] and language agents[10].

Recently, advanced LLMs incorporating reasoning processes have shown superior performance across various tasks[11,12]. However, most of these reasoning models are commercial and subscription-based services. In addition to cost considerations, the handling of highly sensitive data such as pre-clinical differentially expressed genes and private gene sequences raises concerns about uploading to commercial platforms where users have limited control over data governance[13]. Consequently, to reduce the cost and address data privacy concerns, recent studies[14,15] have turned to fine-tuning the open-source LLMs using reinforcement learning policies like online direct preference optimization (DPO)[16] and group relative policy optimization (GRPO)[17], so that these systems can be deployed locally. Nevertheless, no studies have yet explored reasoning-based solutions specifically for the GSA task. In addition, despite evidence that domain-specific knowledge is crucial for the effective LLM fine-tuning[18,19], most of current fine-tuned LLMs is designed and trained for general purpose, relying on general-domain data collected from the internet, which limits their effectiveness in specialty applications such as biomedicine.

---

[*] This research is supported by the NIH Intramural Research Program, National Library of Medicine (NLM)
[†] These authors contributed equally to this work
[‡] The corresponding author

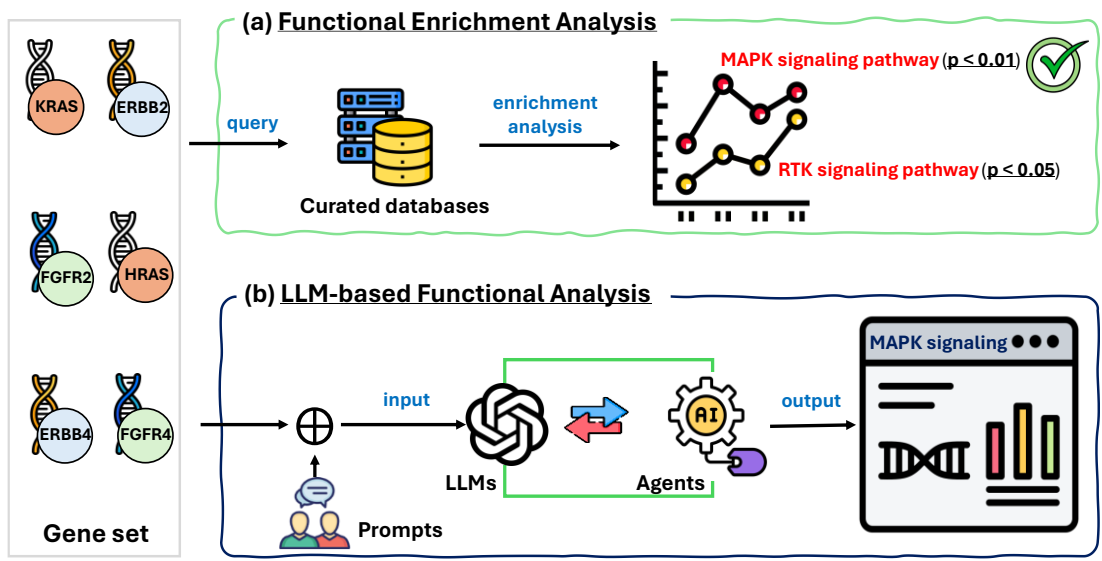

Fig. 1. Example illustration of functional enrichment analysis (a) and LLM-based solutions (b) for the GSA task.

To address this shortcoming, we propose **Gene-R1**, a data-augmented fine-tuning framework that endows lightweight, open-source LLMs with step-by-step reasoning capabilities tailored to the GSA task, aiming to close the performance gap with commercial reasoning LLMs. As illustrated in **Fig.2**, Gene-R1 comprises three modules: knowledge warm-up (KW), reasoning activation (RA), and task alignment (TA). The KW module augments the backbone model with curated knowledge via the pre-training strategy. Using this warmed-up model as the student model, the RA module instills step-by-step inference capabilities by fine-tuning on supervised reasoning examples distilled from the teacher model. Finally, the TA module employs GRPO as the reinforcement learning policy to ensure robust performance in both accuracy and output preferences, with rewards reflecting soft matching to gold-standard labels and strict alignment of outputs and reasoning processes.

We evaluate Gene-R1 in two scenarios: on gene sets whose gold-standard label distributions match those of the fine-tuning data in the RA module (*in-distribution*); on gene sets drawn from different distributions (*out-of-distribution*). In the in-distribution evaluation, Gene-R1 outperforms all comparison methods, demonstrating the effectiveness of the proposed training strategy in instilling step-by-step reasoning capabilities into lightweight LLMs for the GSA task. Meanwhile, in the out-of-distribution evaluation, Gene-R1 matches the performance of both commercial LLMs and large-scale models, underscoring its robust generalizability across diverse gene-set sources.

Overall, our contributions are summarized as follows: (1) We introduce Gene-R1, the first attempt to empower lightweight LLMs with step-by-step reasoning capabilities for the gene set analysis task through data-augmented fine-tuning, which closes the performance gap between open-source and best-performing commercial LLMs. (2) We demonstrate the benefit of priming backbone LLMs with curated domain-specific knowledge for gene set analysis. (3) We validate the robust generalization of Gene-R1 by evaluating it on gene sets from multiple biological sources and across different lightweight LLM variants.

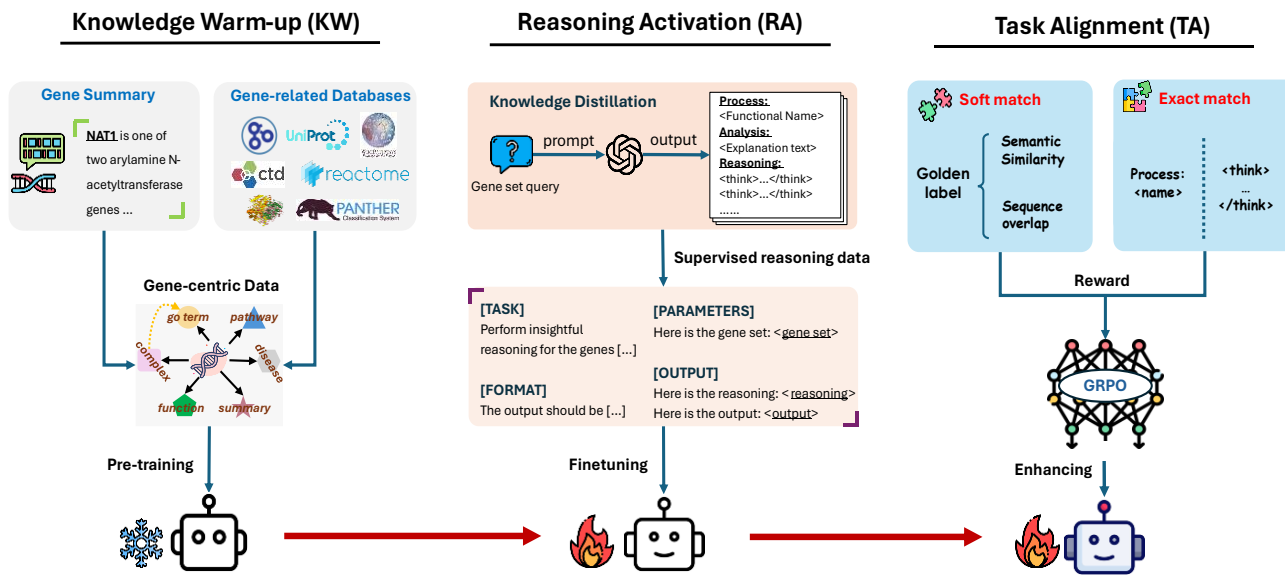

Fig. 2. The overall framework of Gene-R1, which consists of three modules designed to fine-tune lightweight Llama models for the GSA task. These modules are knowledge warm-up, reasoning activation, and task alignment. These modules play a systematic role for incorporating curated knowledge, enhancing reasoning capabilities, and refining the output format of the fine-tuned model, respectively.

## 2. Related Works

Our study primarily intersects gene set analysis and the fine-tuning of large language models.

**Gene set analysis** is a foundational computational approach in bioinformatics that interprets gene expression data by identifying coordinated changes within the predefined groups of genes. Classical methods, most notably Gene Set Enrichment Analysis (GSEA)[5], compare the expression levels of gene sets against curated biological functions documented in the specialized databases. Various tools have been developed for GSEA: g:Profiler[6] performs functional profiling by mapping genes to curated functional resources and detecting statistically significant enrichments, while Enrichr[20] provides a search engine and extensive libraries of annotated gene sets.

Recently, large language models have become valuable tools for GSA, owing to their powerful ability to capture biological context and generate detailed explanations. Jin *et al.*[21] introduced GeneGPT to tackle genomic-related question-answering tasks by augmenting LLMs with external bioinformatics tools. Wu *et al.*[22] presented AutoGen that allows users to build LLM applications like genomic question-answering by composing multiple agents. Hu *et al.*[2] benchmarked five LLMs for the GSA task using prompt engineering and few-shot learning. Wang *et al.*[10] developed the first-of-its-kind AI agent for the same task, which employs multiple domain-specific databases to self-verify the raw outputs of an LLM. In addition, the SPINDOCTOR[9] presents another direction that exploits the summarization capabilities of LLMs to extract biologically plausible processes from gene-function narratives.

**Fine-tuning LLMs** has been a promising way to boost the performance of LLMs on specialized tasks, overcoming the generalization limits of off-the-shelf models. Two efficient approaches are widely used: supervised fine-tuning (SFT) and reinforcement learning (RL).

In SFT, approaches such as MedAlpaca[23] show that modest amounts of curated biomedical question-answering pairs can align general LLMs with domain-specific tasks, improving factual grounding and response style. ClinicalCamel[24] scales this paradigm to more than 1M instructions annotated by clinicians, achieving competitive results on multiple biomedical question-answering datasets. Meanwhile, the RL strategies can further refine LLMs by optimizing generation quality to match domain-expert preferences. For instance, Med-PaLM 2[25] combines instruction tuning with physician-provided reward models, surpassing 85% answer accuracy on USMLE-style exams. DeepSeek[11] employs the GRPO policy to finetune the LLMs and obtain the state-of-the-art performance across multiple tasks. In the biomedical domain, UltraMedical[26] collections facilitate the fine-tuning of several advanced medical LLMs based on the Llama-3 series. Recently, cell-o1[27] trained a 7B-parameter LLM for the CellPuzzles task by equipping the RL with batch-level rewards.

## 3. Methodology

As illustrated in **Fig. 2**, Gene-R1 comprises three cascaded steps: knowledge warm-up, reasoning activation, and task alignment. Each of them is designed to respectively incorporate prior gene-relevant knowledge, learn biological reasoning patterns, and match format-specific preferences.

The knowledge warm-up module aims to equip the model with a foundational understanding of gene symbols and basic biological terminology. Most existing open-source LLMs are fine-tuned on general-domain corpora so that they are rarely exposed to domain-specific knowledge such as gene annotations. As a result, without prior exposure or contextual grounding, gene symbols are typically treated as meaningless strings by these LLMs and are likely to induce hallucination.

We hypothesize that LLMs can significantly benefit from early-stage exposure to gene-relevant knowledge, and therefore, we collect gene knowledge from several databases (*e.g.*, GO, CTD[28], UniProtKB[29], *etc.*) and consolidate them into a gene-centric relational dataset $\{d_i\}_{i=1}^N$ as shown in **Tab.1**, which enables the model to learn meaningful associations between gene symbols and their biological contexts. Based on this dataset, we leverage **pre-training** to incorporate the domain-specific knowledge. Each instance $d_i$ is considered as a declarative sentence to optimize the parameter set of the backbone LLMs by minimizing the token-level likelihood probabilities. After pre-training, the model is better able to recognize the gene symbol rather than hallucinating a factually incorrect definition. For example, the base model recognizes GHRHR as "*a receptor that regulates the expression of ERAD-related genes*", whereas the pre-trained model can correctly recall gene knowledges and output "*The gene GHRHR is involved in growth hormone signaling and growth regulation.*"

Table 1. Statistics of data using for the Gene-R1 training.

| | # Source | # Instance | Description |
|---|---|---|---|
| KW via model pre-training | 8 | 244,754 | Gene-centric relational data from GO, UniProtKB, CTD, Reactome[30], Wikipathway[31], Panther[32], COURM[33], and NCBI[34] |
| RA via model fine-tuning | 3 | 9,873 | Gene sets with reasoning process generated by GPT-o1. These gene sets are sampled from GO:BP, GO:MF, and GO:CC. |
| TA via reinforcement learning | 5 | 13,327 | Gene sets with ground-truth labels from source databases: GO, Omics analysis platform[35], PubMed, Reactome, and MSigDB |

Building upon the pre-trained model , the reasoning activation module further enables the model to perform reasoning with gene utilities and associations.

Due to the lack of existing reasoning annotations and the impracticality of manually curating step-by-step reasoning processes for gene sets, we turn to employing the GPT-o1 model to generate reasoning chains for gene sets collected from the three branches of the GO database. Specifically, we build a supervised reasoning corpus $\{q_i, r_i, f_i\}_{i=1}^K$ for model **fine-tuning** as shown in **Tab.1**, where each instance includes a gene-set query ($q$), a step-by-step reasoning process ($r$), and the generated biological function ($f$). The prompt template used for GPT-o1 to generate such instances is provided in **Tab.2**. Importantly, to ensure high-quality supervised data, we only retain those instances in which the generated function achieves a similar score greater than 0.7 with the ground-truth label. The resulting dataset is then used to fine-tune the pre-trained model, enabling it to learn biologically grounded reasoning patterns of gene sets. After fine-tuning the model on the supervised reasoning corpus, the model could make biological inference steps for the input gene set such as "*the CRY2 gene functions in the regulation of gene expression in response to cellular stress, which impacts the regulation of phosphate homeostasis.*"

Table 2. The template for the GPT-o1 model to generate reasoning corpora given gene sets.

**System:** You are an efficient and insightful assistant to a molecular biologist.
**User:** Perform insightful reasoning for the interacting proteins and write a critical analysis of the biological functions based on your reasoning.
Propose a brief name for the prominent biological functions performed by the system, such as biological process, molecular function, cellular component, and so on.
The proposed name and critical analysis should:
Be concise; avoid unnecessary words.
Be textual; do not use format symbols such as "*", "-", or other tokens.
Be specific; avoid overly general statements such as "the proteins are involved in various cellular processes."
Be factual; do not editorialize.
For each reasoning point and the critical analysis, describe the supporting information. They should:
Be comprehensive; collect various gene functions from different aspects, including gene summaries, enrichment analysis, gene complexes, gene domains, pathway analysis, and more.
Be complete; ensure no necessary or helpful genes in the given gene set are missed.
Be convincing; do not generate ambiguous statements for any genes.
Be ample; provide long, high-quality, and credible evidence for the proposed process name.
Here is the gene set: **{genes}**
The analysis must include the following format:
1. Put the name at the top of the analysis as **"*Process: <name>*"**.
2. The reasoning process must be placed at the bottom of the analysis, starting with the message: "Reasoning: ".
3. Each reasoning step should be organized within the **"<think></think>"** tags.

Finally, we curate a new benchmark dataset $\{g_i, a_i\} \mid_{i=1}^{M}$ containing the gene sets $(g)$ and their ground-truth functional annotations $(a)$, as shown in **Tab.1**, to train a **reinforcement learning** algorithm, aiming to enhance the accuracy of the fine-tuned model.

In practice, we introduce the task alignment module equipped with the effective reward functions to guide the fine-tuned model's reasoning process $r_{g_i}$ and its generated biological function name $f_{g_i}$ for the gene set $g_i$ toward the correct annotation $a_i$. This component boosts both the accuracy and output formatting via the GRPO policy. Specifically, to improve accuracy, we implement a soft-match reward function that captures both semantic similarity and sequence overlap between $f_{g_i}$ and $a_i$, which integrates the MedCPT[36] score for semantic similarity and the longest common subsequence (LSC) score[37] for lexical alignment. In parallel, to ensure the output adheres to the expert-preference format that includes a "*Process*" identifier for the prominent biological function and uses a "$<think></think>$" tag pair to separate different reasoning steps, we retain the exact-match reward function used in the original GRPO framework. This dual-reward strategy allows the model to optimize both the semantic correctness and structural compliance.

## 4. Experiments

Our experiments are designed to answer the following questions:

Q1. How does the Gene-R1 perform compared with both the state-of-the-art open-source and commercial LLMs?

Q2. How well does Gene-R1 generalize across different gene sets (in-distribution *vs*. out-of-distribution evaluation)?

Q3. What are the contributions of each individual module within Gene-R1?

### 4.1. *Datasets*

For model evaluation, we curated five independent datasets from the GO database, proteomics analysis, and a molecular function database repository. We made sure none of the test data was previously used during the development phase of Gene-R1. As shown in **Tab.3**, all GO datasets are employed for in-distribution evaluation, while the other two datasets are reserved for out-of-distribution assessment.

Table 3. Statistics of data used for the Gene-R1 evaluation.

|  | # Sets | #Genes in a Set | Avg. #Genes | Source |
|---|---|---|---|---|
| GO:BP | 1,000 | 3 to 456 | 48.3 | Literature curation |
| GO:MF | 340 | 1 to 5,973 | 78.2 | Literature curation |
| GO:CC | 168 | 2 to 13,075 | 249.5 | Literature curation |
| NeST | 50 | 5 to 323 | 2.2 | Proteomics analysis[35] |
| MsigDB | 56 | 4 to 200 | 3.0 | Molecular function |

### 4.2. *Experimental Setting*

**Task definition:** Given a group of $k$ genes, i.e., a gene set $(\mathcal{S} = \{g_i \mid_{i=1}^{k}\})$, our goal is applying the fine-tuned LLM $(\mathcal{L})$ on $\mathcal{S}$ to generate the biologically plausible function $(\mathcal{F})$ and coherent explanatory context $(\mathbb{C})$ through a step-by-step reasoning process.

**Evaluation metrics:** To comprehensively evaluate the biological plausibility of the functional names generated by Gene-R1 with the gold-standard labels, we employed two complementary metrics: **ROUGE** (Recall-Oriented Understudy for Gisting Evaluation)[37] to measure the lexical overlap and **Similarity Score** to quantify the semantic relevance. To mitigate potential bias inherent to a single semantic encoder, we calculated semantic similarity using three biomedical encoders— MedCPT, SentenceBERT[38], and SapBERT[39]—and reported the average scores.

**Implementations:** All training and evaluation are implemented with python 3.13.5 and torch 2.7.1 on the AWS (Amazon Web Services) services (8 GPU cards). Other required software packages are transformers (4.53.2), trl (0.19.0), accelerate (1.8.1), deepspeed (0.17.2).

### 4.3. *Backbone and comparison LLMs*

After investigating criteria involving accessibility, fine-tuning cost, performance, and general usability, we select the Meta Llama[40] as the backbone LLMs for Gene-R1. Specifically, we use Llama3.1 (8B parameters) and Llama3.2 (1B and 3B parameters) to demonstrate the flexibility of proposed fine-tuning pipeline. Furthermore, we compare Gene-R1 with the aligned Llama models and widely used commercial GPT models: GPT-4, GPT-4o, o1, and o3-mini. All Llama models are accessed from the hugging-face community, while the GPT models are provided by the Azure API.

## 5. Results

### 5.1. *In-distribution Evaluation*

To address Q1 that is related to the accuracy assessment of Gene-R1, we evaluate a range of LLMs on 1,508 gene sets whose label distributions matched those of the fine-tuning data used in the RA module.

As shown in **Tab.4**, Gene-R1 consistently outperforms all baselines on both ROUGE and semantic-similarity metrics. Specifically, compared to Llama3 models, Gene-R1 increased ROUGE scores by 133.1%, 133.9%, and 257.3% in average, respectively. Compared to the GPT series reasoning models, it improved these metrics by 82.0%, 77.5%, and 192.8% in average. The marked gain in ROUGE-2 indicates that the biological function names predicted by Gene-R1 exhibit substantially longer n-gram overlaps with the gold-standard labels, demonstrating that our fine-tuning workflow effectively captures distributional patterns for precise sequence generation. This high ROUGE performance also translates into superior semantic alignment. By averaging similarity scores from MedCPT, SentenceBERT, and SapBERT, we found that Gene-R1 achieved gains of 5.9%, 16.9%, and 18.3% respectively over Llama-based models and 2.9%, 8.4%, and 6.1% respectively over GPT-based models across the three evaluation datasets. These results confirm that Gene-R1 closes the performance gap with commercial LLMs.

Additionally, the paired comparisons between the backbone models and Gene-R1 at the 1B, 3B, and 8B parameter scales show that our fine-tuning workflow preserves strong performance across Llama variants of different sizes. Specifically, Gene-R1 consistently delivers over 15% gains in terms of the semantic similarity scores on every evaluation dataset, underscoring its robustness. This reliability opens new avenues for applying Gene-R1 to a wide range of tunable and open-source LLMs for the specificized downstream task, thereby being able to shorten development cycles.

Table 4. Performance of Gene-R1 on gene sets derived from three branches of the GO database (In-distribution Evaluation). **R.-(*)** denotes the ROUGE score under different metrics. **Score (avg.)** represents the semantic similarity score averaged across MedCPT, SentenceBERT, and SapBERT. The best results for each dataset are highlighted in **bold**. $\Delta$ denotes the relative improvement calculated as $(\mathbf{bold} - x)/x * 100\%$. The improvements are significant (p-value $< 0.05$) according to a two-tailed paired t-test at a 95% confidence interval.

| | Models | R.-L | $\Delta$ | R.-1 | $\Delta$ | R.-2 | $\Delta$ | Score (avg.) | $\Delta$ |
|---|---|---|---|---|---|---|---|---|---|
| GO: BP | Llama3-1B | 0.107 | 158.9% | 0.109 | 165.1% | 0.049 | 122.4% | 0.468 | 38.2% |
| | Llama3-3B | 0.083 | 233.7% | 0.087 | 232.2% | 0.011 | 890.9% | 0.505 | 28.1% |
| | Llama3-8B | 0.133 | 108.3% | 0.146 | 97.9% | 0.025 | 336.0% | 0.562 | 15.1% |
| | Llama3-70B | 0.196 | 41.3% | 0.212 | 36.3% | 0.062 | 75.8% | 0.611 | 5.9% |
| | GPT-4 | 0.184 | 50.5% | 0.201 | 43.8% | 0.049 | 122.4% | 0.614 | 5.4% |
| | GPT-4o | 0.184 | 50.5% | 0.207 | 39.6% | 0.036 | 202.8% | 0.629 | 2.9% |
| | o1 | 0.164 | 68.9% | 0.178 | 62.4% | 0.040 | 172.5% | 0.626 | 3.4% |
| | o3-mini | 0.154 | 79.9% | 0.167 | 73.1% | 0.033 | 230.3% | 0.614 | 5.4% |
| | Gene-R1(1B) | 0.225 | 23.1% | 0.232 | 24.6% | 0.075 | 45.3% | 0.617 | 4.9% |
| | Gene-R1(3B) | 0.229 | 21.0% | 0.238 | 21.4% | 0.080 | 36.3% | 0.623 | 3.9% |
| | Gene-R1(8B) | **0.277** | / | **0.289** | / | **0.109** | / | **0.647** | / |
| GO: MF | Llama3-1B | 0.028 | 1057.1% | 0.028 | 1067.9% | 0.002 | 7900.0% | 0.470 | 45.7% |
| | Llama3-3B | 0.095 | 241.1% | 0.095 | 244.2% | 0.022 | 627.3% | 0.538 | 27.3% |
| | Llama3-8B | 0.099 | 227.3% | 0.102 | 220.6% | 0.027 | 492.6% | 0.564 | 21.5% |
| | Llama3-70B | 0.114 | 184.2% | 0.114 | 186.8% | 0.024 | 566.7% | 0.586 | 16.9% |
| | GPT-4 | 0.100 | 224.0% | 0.101 | 223.8% | 0.023 | 595.7% | 0.588 | 16.5% |
| | GPT-4o | 0.096 | 237.5% | 0.097 | 237.1% | 0.020 | 700.0% | 0.584 | 17.3% |
| | o1 | 0.147 | 120.4% | 0.150 | 118.0% | 0.037 | 332.4% | 0.632 | 8.4% |
| | o3-mini | 0.119 | 172.3% | 0.120 | 172.5% | 0.029 | 451.7% | 0.611 | 12.1% |
| | Gene-R1(1B) | 0.315 | 2.9% | 0.318 | 2.8% | 0.122 | 31.1% | 0.661 | 3.6% |
| | Gene-R1(3B) | 0.316 | 2.5% | 0.319 | 2.5% | 0.122 | 31.1% | 0.661 | 3.6% |
| | Gene-R1(8B) | **0.324** | / | **0.327** | / | **0.160** | / | **0.685** | / |
| GO: CC | Llama3-1B | 0.039 | 546.2% | 0.039 | 564.1% | 0.008 | 962.5% | 0.449 | 43.9% |
| | Llama3-3B | 0.075 | 236.0% | 0.076 | 240.8% | 0.013 | 553.8% | 0.497 | 30.0% |
| | Llama3-8B | 0.092 | 173.9% | 0.093 | 178.5% | 0.028 | 203.6% | 0.536 | 20.5% |
| | Llama3-70B | 0.091 | 176.9% | 0.091 | 184.6% | 0.022 | 286.4% | 0.546 | 18.3% |
| | GPT-4 | 0.093 | 171.0% | 0.092 | 181.5% | 0.016 | 431.3% | 0.562 | 14.9% |
| | GPT-4o | 0.105 | 140.0% | 0.105 | 146.7% | 0.020 | 325.0% | 0.573 | 12.7% |
| | o1 | 0.144 | 75.0% | 0.148 | 75.0% | 0.038 | 123.7% | 0.609 | 6.1% |
| | o3-mini | 0.139 | 81.3% | 0.142 | 82.4% | 0.031 | 174.2% | 0.598 | 8.0% |
| | Gene-R1(1B) | 0.222 | 13.5% | 0.226 | 14.6% | 0.054 | 57.4% | 0.618 | 4.5% |
| | Gene-R1(3B) | 0.143 | 76.2% | 0.143 | 81.1% | 0.035 | 142.9% | 0.578 | 11.8% |
| | Gene-R1(8B) | **0.252** | / | **0.259** | / | **0.085** | / | **0.646** | / |

## 5.2. *Out-of-distribution Evaluation*

To address Q2, which concerns the generalization of Gene-R1, we evaluated its performance on 106 gene sets curated by Hu *et al.*[2] and filtered by Wang *et al.*[10] These gene sets are associated with gold-standard labels that exhibit distributional characteristics distinct from those used during model fine-tuning. For example, many labels of gene sets in these two datasets are abbreviated (e.g., "*TNFR signaling*"), whereas the corresponding in-distribution annotations are more informative and descriptive (e.g., "*regulation of the tumor necrosis factor receptor signaling*").

As shown in **Tab.5**, although Gene-R1 does not outperform all comparison methods on every evaluation metric, it is consistently comparable to the best baselines across both datasets. Notably, significance tests on similarity scores between Gene-R1 and the top-performing LLMs reveal no significant differences, indicating that Gene-R1 achieves performance on par with both commercial LLMs such as GPT-4 and large-scale models like Llama3.3-70B. It is also worth mentioning that Gene-R1 consistently surpasses the o1 and o3-mini, demonstrating that the task-specific reasoning generated by Gene-R1 better aligns with biological functions than general-purpose reasoning LLMs. These findings highlight the strong generalization capability of Gene-R1 for diverse gene sets.

Table 5. Performance comparison between Gene-R1 and other LLMs on out-of-distribution datasets. The best results in different datasets are **bold**. "**n.s.**" denotes no significant difference (p-value > 0.05) according to a two-tailed paired t-test at the 95 % confidence level.

| Datasets | Models | ROUGE-L | ROUGE-1 | ROUGE-2 | Similarity Score (avg.) |
|---|---|---|---|---|---|
| NeST | Llama3-1B | 0.149 | 0.154 | 0.056 | 0.522 |
| | Llama3-3B | 0.152 | 0.162 | 0.033 | 0.570 |
| | Llama3-8B | 0.197 | 0.210 | 0.073 | 0.610 |
| | Llama3-70B | 0.220 | 0.234 | 0.071 | 0.633 (n.s.) |
| | GPT-4 | 0.239 | 0.252 | 0.082 | **0.638 (n.s.)** |
| | GPT-4o | 0.185 | 0.200 | 0.065 | 0.611 |
| | o1 | 0.153 | 0.156 | 0.028 | 0.618 |
| | o3-mini | 0.179 | 0.190 | 0.035 | 0.625 |
| | Gene-R1 (1B) | **0.249** | **0.252** | 0.071 | 0.630 |
| | Gene-R1 (3B) | 0.238 | 0.243 | **0.091** | 0.635 |
| | Gene-R1 (8B) | 0.216 | 0.223 | 0.089 | 0.616 |
| MsigDB | Llama3-1B | 0.033 | 0.033 | 0.005 | 0.463 |
| | Llama3-3B | 0.164 | 0.164 | 0.030 | 0.563 |
| | Llama3-8B | 0.177 | 0.177 | 0.037 | 0.596 |
| | Llama3-70B | 0.195 | 0.195 | 0.070 | 0.611 |
| | GPT-4 | **0.239** | **0.239** | 0.074 | 0.628 (n.s.) |
| | GPT-4o | 0.220 | 0.220 | 0.046 | **0.632 (n.s.)** |
| | o1 | 0.167 | 0.167 | 0.031 | 0.625 |
| | o3-mini | 0.165 | 0.165 | 0.011 | 0.605 |
| | Gene-R1 (1B) | 0.177 | 0.177 | 0.041 | 0.605 |
| | Gene-R1 (3B) | 0.214 | 0.218 | **0.077** | 0.622 |
| | Gene-R1 (8B) | 0.203 | 0.203 | 0.068 | 0.625 |

## 5.3. *Ablation Experiments*

Additional ablation experiments were conducted to address Q3 by investigating the individual contributions of each module within Gene-R1.

We designed an incremental fine-tuning setup in which modules are introduced one at a time, enabling us to quantify their independent effects on model performance. Specifically, we produced two intermediate variants of Gene-R1: one using only the gene-centric relational data from the knowledge warm-up module (i.e., w/ KW) and another with both realtional data and the supervised reasoning data from the reasoning activation modue (i.e., w/ KW&RA). To ensure robust evaluation, we applied different variants acorss multiple backbone models (Llama 1B, 3B, and 8B) and datasets (in-distribution and out-of-distribution). The results are summarized in **Fig.3**.

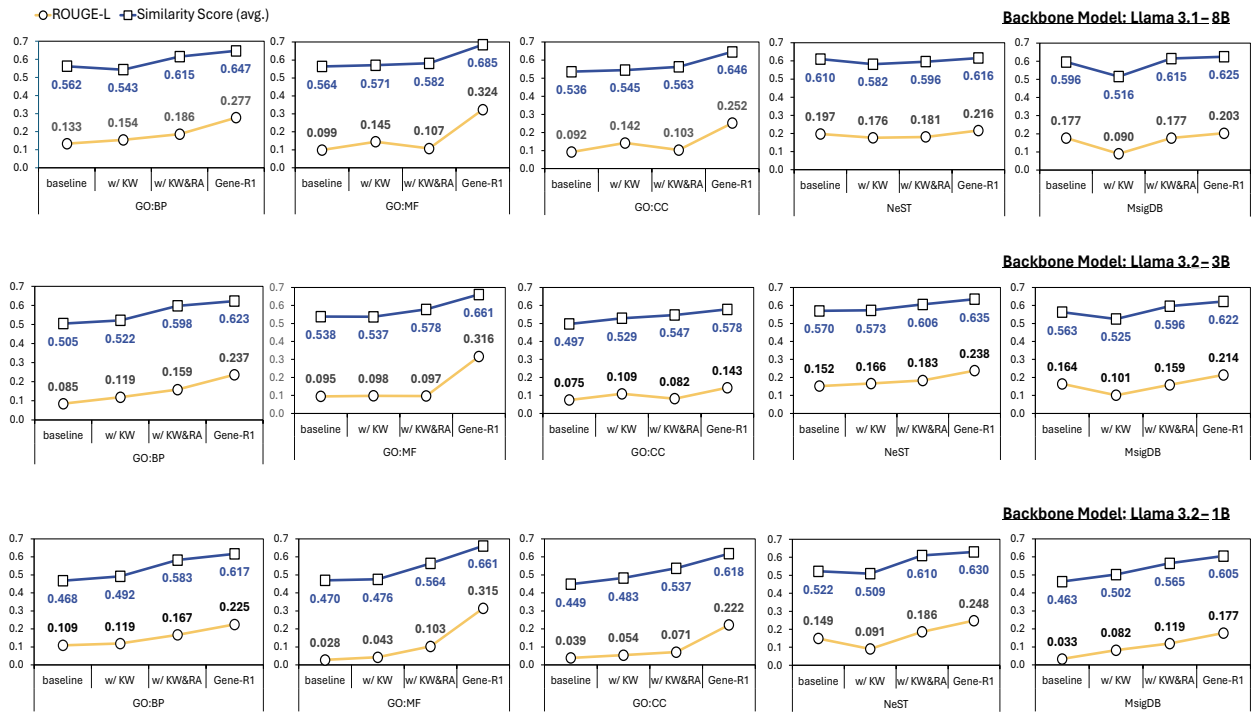

Fig. 3. Performance comparison of the individual module contributes to Gene-R1. The experiments were conducted using Llama with 1B, 3B, and 8B parameters, respectively. The y-axis represents the performance score, while the x-axis indicates the sequential addition of modules to the baseline backbone model.

On the in-distribution datasets (i.e., GO:BP, GO:MF, and GO:CC), each module demonstrated a clear and consistent positive contribution to the performance of Gene-R1 compared to the baseline model. In constract, the results on the out-of-distribution datasets (i.e., NeST and MsigDB) revealed that no individual module consistency improved performance in solation. Instead, the combination of all three modules (i.e., w/ KW&RA&TA) provided systematic improvement. Notably, the most significant performance gain was observed when the TA module was incorporated, underscoring the critical role of reinforcement learning in enhancing the effectiveness of Gene-R1. For example, on the GO:MF benchmark, the model using Llama3.1-8B backbone achieved a 10.3 percentage point increase in semantic similarity and a 0.217 absolute gain in ROUGE-L.

## 6. Discussion

**Advantages of data augmentation for lightweight LLM fine-tuning.** In this work, we present Gene-R1, an effective pipeline that enables open-source LLMs with fewer parameters to achieve performance comparable to larger models, including popular commercial alternatives. The high cost and opaque nature of commercial LLMs raise concerns including budget and data privacy, which hinder their deployment in real world settings. Although recent fine-tuned LLMs have shown promise in domain-specific tasks, they often generate fabricated content such as incorrect definitions of technical terms, due to limited exposure to specialized knowledge. In contrast, Gene-R1 incorporates domain knowledge to its fine-tuning workflow that uses accessible lightweight open-source LLMs. This allows local deployment with limited computational resources and can reduce domain-specific hallucinations and inference expenditure.

**Different reward functions for reinforcement learning**. Reinforcement learning is crucial for Gene-R1's performance. To further investigate this component, we examined how the choice of reward modeling and reward function design impacts the performance of Gene-R1.

We first compare the GRPO with standard online DPO, where the reward model is trained on pre-generated responses[a]. Then, we relax the reward formulation by introducing a "soft-match" component that assigns partial credit for outputs that are semantically related to the gold-standard labels. For this evaluation, we use the most stable model (Llama 8B) and the GO:BP bechmark as shown in Fig.3. As summarized in **Fig.4 (a)**, the GRPO with soft-match rewards consistently outperforms both the online DPO and GRPO with only exact-match rewards.

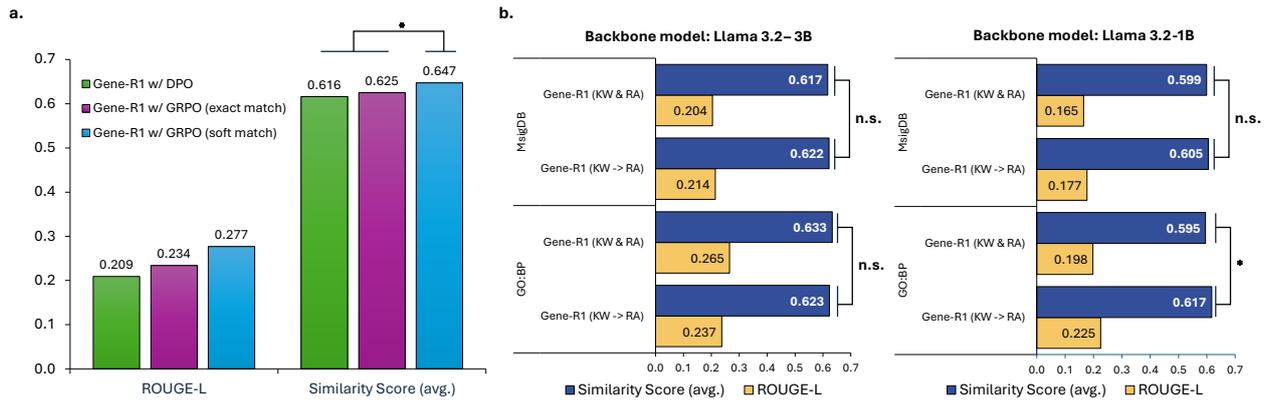

Fig. 4. Alation experiments for Gene-R1. (a) The comparison for Gene-R1 with different reinforcement learning settings. (b) The results of different approaches for the incorporation of gene-centric relational data. "KW & RA" denotes merged strategy, while "KW -> RA" denote the cascade strategy. "*" indicates the significant improvement (p-value < 0.05) according to a two-tailed paired t-test at a 95% confidence interval, while "**n.s.**" indicates not significant.

**Different strategies to incorporate domain-specific knowledge**. This work has shown that domain-specific knowledge provides the foundational biological context necessary for Gene-R1's step-by-step reasoning. To explore the optimal strategy for injecting the gene-centric relational knowledge into model training, we investigated two separate approaches: the *Cascade* strategy (employed in Gene-R1) and a *Merged* strategy.

In the alternative merged approach, relational data are appended directly to the reasoning exemplars and introduced solely during the RA stage. This allows the model to learn structural knowledge and reasoning patterns simultaneously, without a dedicated pre-training phase. We evaluate them on two representative benchmarks containing the largest number of gene sets: GO:BP (in-distribution) and MsigDB (out-of-distribution). The results shown in **Fig.4 (b)**, demonstrating that the cascade strategy yields only modest improvements over the merged approach in terms of both similarity score; however, most of these differences are not statistically significant.

---

[a] In Online DPO, a reward function is used to determine the chosen and rejected response during training. In our work, we train the reward function using pre-generated model responses from GPT. Specifically, for each gene set query, we generate four types of responses in preference order: 1. Response with correct answer and contains reasoning; 2. Response with correct answer and does not contain reasoning; 3. Response with incorrect answer and contains reasoning; 4. Response with incorrect answer and does not contain reasoning. These responses are then used to create a pair-wise dataset consisting of chosen-rejected pairs. Given two responses, the reward model is trained to distinguish and select the better response following the preference order.

**Error analysis**. In addition to being constrained by the inherent limitations of the Llama model for the gene set analysis task, the primary source of errors in Gene-R1 comes from its informal reasoning processes. As shown in **Tab.6**, the models fail to consistently achieve 100% accuracy in output formatting. Some outputs either lack a clearly defined biological function name or fail to adhere to a valid step-by-step reasoning structure, ultimately resulting in incorrect predictions. A promising solution to this shortcoming is to manually create a subset of high-quality reasoning annotations to guide the teacher policy (e.g., the GPT-o1 model) in generating better supervised data. Alternatively, customized reward functions can be introduced to encourage the production of longer and more coherent reasoning chains.

Table 6. The proportion of correct output format generated by Gene-R1

|         | **Gene-R1 (1B)**  | **Gene-R1 (3B)**  | **Gene-R1 (8B)**  |
|---------|-------------------|-------------------|-------------------|
| GO:BP   | 95.6% (956/1000)  | 42.4% (424/1000)  | 93.8% (938/1000)  |
| GO:MF   | 94.4% (321/340)   | 94.4% (321/340)   | 85.0% (289/340)   |
| GO:CC   | 91.7% (154/168)   | 42.9% (72/168)    | 88.1% (148/168)   |
| NeST    | 96.0% (48/50)     | 100% (50/50)      | 100% (50/50)      |
| MsigDB  | 91.1% (51/56)     | 100% (56/56)      | 100% (56/56)      |

**Limitations**. Despite these advancements achieved by Gene-R1, it currently relies on manual data collection for the training workflow, which limits the flexibility when applied to novel or underrepresented genes. Moreover, Gene-R1 may still hallucinate plausible-sounding but incorrect functions when operating on unseen genes which are outside its training domain. Furthermore, our evaluation primarily focuses on biological function annotation, leaving the model's transferability to other ontologies (e.g., disease ontology, phenotype ontology) as open questions.

## 7. Conclusions

In this study, we present Gene-R1 to equip lightweight, open-source LLMs with strong reasoning capabilities specifized for the gene set analysis task, effectively narrowing the performance gap with proprietary, large-scale models. The effectiveness of Gene-R1 is shown by both the in-distribution and out-of-distribution evaluations on five datasets containing 1604 gene sets. LLMs trained with general domain data are highly capable of linguistic tasks, knowledge recalling and reasoning, but they often fail in domain specific tasks. We believe one best way to fully utilize the power of the LLMs is through incoporating domain knowledges, and methods like our Gene-R1 can enable powerful AIs for wider and more diffcult tasks.

## References

1. de Leeuw, C., Neale, B., Heskes, T. *et al.* The statistical properties of gene-set analysis. *Nat Rev Genet* **17**, 353–364 (2016).
2. Hu, M., Alkhairy, S., Lee, I. *et al.* Evaluation of large language models for discovery of gene set function. *Nat Methods* **22**, 82–91 (2025).

3. Ma, K., Huang, S., Ng, K.K. *et al*. Saturation mutagenesis-re-inforced functional assays for disease-related genes. Cell **187**, 6707–6724.e22 (2024).

4. Tachmazidou, I., Hatzikotoulas, K., Southam, L. *et al*. Identification of new therapeutic targets for osteoarthritis through genome-wide analyses of UK Biobank data. *Nat Genet* **51**, 230–236 (2019).

5. Subramanian, A., Tamayo, P., Mootha, V.K. *et al*. Gene set enrichment analysis: a knowledge-based approach for interpreting genome-wide expression profiles. *Proceedings of the National Academy of Sciences* **102**, 15545-15550 (2005).

6. Raudvere, U., Kolberg, L., Kuzmin, I. *et al*. g: Profiler: a web server for functional enrichment analysis and conversions of gene lists (2019 update). *Nucleic acids research* **47**, W191-W198 (2019).

7. Ashburner, M., Ball, C., Blake, J. *et al*. Gene Ontology: tool for the unification of biology. *Nat Genet* **25**, 25–29 (2000).

8. Liberzon, A., Birger, C., Thorvaldsdóttir, H. *et al*. The molecular signatures database hallmark gene set collection. *Cell systems* **1**, 417-425 (2015).

9. Joachimiak, M.P., Caufield, J.H., Harris, N.L. *et al*. Gene set summarization using large language models. *ArXiv*, arXiv-**2305** (2024).

10. Wang, Z., Jin, Q., Wei, CH. *et al*. GeneAgent: self-verification language agent for gene-set analysis using domain databases. *Nat Methods* 1-9 (2025).

11. Liu, A., Feng, B., Xue, B. *et al*. Deepseek-v3 technical report. *arXiv*:**2412**.19437 (2024).

12. Jaech, A., Kalai, A., Lerer, A. *et al*. Openai o1 system card. *arXiv*:**2412**.16720 (2024).

13. Yang, Y., Jin, Q., Leaman, R. *et al*. Ensuring safety and trust: Analyzing the risks of large language models in medicine. *arXiv* preprint arXiv:**2411**.14487 (2024).

14. Wang, Z., Du, Y., Sun, Z. *et al*. Re2llm: reflective reinforcement large language model for session-based recommendation. *Proceedings of the AAAI Conference on Artificial Intelligence* **39**, 12827-12835 (2025).

15. Trung, L., Zhang, X., Jie, Z. *et al*. Reft: Reasoning with reinforced fine-tuning. *Proceedings of the 62nd Annual Meeting of the Association for Computational Linguistics* **1**, 7601-7614 (2024).

16. Qi, B., Li, P., Li, F. *et al*. Online dpo: Online direct preference optimization with fast-slow chasing. *arXiv* preprint arXiv:**2406**.05534 (2024).

17. Shao, Z., Wang, P., Zhu, Q. *et al*. Deepseekmath: Pushing the limits of mathematical reasoning in open language models. *arXiv* preprint arXiv:**2402**.03300 (2024).

18. Xue, Z., Li, L., Tian, S. *et al*. Domain knowledge is all you need: A field deployment of llm-powered test case generation in fintech domain. *Proceedings of the 2024 IEEE/ACM 46th International Conference on Software Engineering: Companion Proceedings* 314-315 (2024).

19. Song, Z., Yan, B., Liu, Y. *et al*. Injecting domain-specific knowledge into large language models: a comprehensive survey. *arXiv* preprint arXiv:**2502**.10708 (2025).

20. Kuleshov, M. V., Jones, M. R., Rouillard, A. D. *et al*. Enrichr: a comprehensive gene set enrichment analysis web server 2016 update. *Nucleic acids research* **44**, W90-W97 (2016)

21. Jin, Q., Yang, Y., Chen, Q., & Lu, Z. Genegpt: Augmenting large language models with domain tools for improved access to biomedical information. *Bioinformatics* **40**, btae075 (2024).

22. Wu, Q., Bansal, G., Zhang, J. *et al*. Autogen: Enabling next-gen LLM applications via multi-agent conversations. *First Conference on Language Modeling* (2024).

23. Han, T., Adams, L.C., Papaioannou, J.M. *et al*. MedAlpaca--an open-source collection of medical conversational AI models and training data. *arXiv* preprint arXiv:**2304**.08247 (2023).

24. Toma, A., Lawler, P.R., Ba, J. *et al*. Clinical camel: An open expert-level medical language model with dialogue-based knowledge encoding. *arXiv* preprint arXiv:**2305**.12031 (2023).

25. Singhal, K., Tu, T., Gottweis, J. *et al.* Toward expert-level medical question answering with large language models. *Nat Med* **31**, 943–950 (2025).

26. Zhang, K., Zeng, S., Hua, E. *et al*. Ultramedical: Building specialized generalists in biomedicine. *Advances in Neural Information Processing Systems* **37**, 26045-26081 (2024).

27. Fang, Y., Jin, Q., Xiong, G. *et al*. Cell-o1: Training LLMs to Solve Single-Cell Reasoning Puzzles with Reinforcement Learning. *arXiv* preprint arXiv:**2506**.02911 (2025).

28. Davis, A.P., Grondin, C.J., Johnson, R.J. *et al*. Comparative toxicogenomics database (CTD): update 2021. *Nucleic acids research* **49**, D1138-D1143 (2021).

29. Coudert, E., Gehant, S., De Castro *et al*. Annotation of biologically relevant ligands in UniProtKB using ChEBI. *Bioinformatics* **39**, btac793 (2023).

30. Fabregat, A., Jupe, S., Matthews, L. et al. The reactome pathway knowledgebase. Nucleic acids research 46, D649-D655 (2018).

31. Agrawal, A., Balcı, H., Hanspers, K. *et al*. WikiPathways 2024: next generation pathway database. *Nucleic acids research* **52**, D679-D689 (2024).

32. Thomas, P.D., Ebert, D., Muruganujan, A. *et al*. PANTHER: Making genome - scale phylogenetics accessible to all. *Protein Science* **31**, 8-22 (2022).

33. Giurgiu, M., Reinhard, J., Brauner, B. *et al*. CORUM: the comprehensive resource of mammalian protein complexes—2019. *Nucleic acids research* **47**, D559-D563 (2019).

34. Geer, L.Y., Marchler-Bauer, A., Geer, R.C. *et al*. The NCBI biosystems database. *Nucleic acids research* **38**, D492-D496 (2010).

35. Zheng, F., Kelly, M.R., Ramms, D.J. *et al*. Interpretation of cancer mutations using a multiscale map of protein systems. *Science* **374**, eabf3067 (2021).

36. Jin, Q., Kim, W., Chen, Q. *et al*. Medcpt: Contrastive pre-trained transformers with large-scale pubmed search logs for zero-shot biomedical information retrieval. *Bioinformatics* **39**, btad651 (2023).

37. Lin, C.Y. Rouge: A package for automatic evaluation of summaries. In Text summarization branches out, 74-81 (2004).

38. Reimers, N., & Gurevych, I. Sentence-BERT: Sentence Embeddings using Siamese BERT-Networks. *Proceedings of the 2019 Conference on Empirical Methods in Natural Language Processing and the 9th International Joint Conference on Natural Language Processing (EMNLP-IJCNLP)*, 3982-3992 (2019).

39. Liu, F., Shareghi, E., Meng, Z. *et al*. Self-alignment pretraining for biomedical entity representations. *North American Association for Computational Linguistics 2021*, *Association for Computational Linguistics* (*ACL*), 4228-4238 (2021).

40. Touvron, H., Lavril, T., Izacard, G. *et al*. Llama: Open and efficient foundation language models. *arXiv* preprint arXiv:**2302**.13971 (2023).